\documentclass[11pt]{article}

\usepackage[utf8]{inputenc}
\usepackage[T1]{fontenc}
\usepackage{times} 
\usepackage{graphicx}
\usepackage{subcaption}
\usepackage{xcolor}
\usepackage{amssymb,amsmath,amsthm}
\usepackage{placeins}
\usepackage{hyperref}
\usepackage{pdfpages}
\usepackage{authblk}

\usepackage[margin=1in]{geometry}

\title{Explorations of Epidemiological Dynamics across Multiple Population Hubs}

\author[1,2]{Daniel Perkins}
\author[1,2]{Davis Hunter}
\author[1,3]{Drake Brown}
\author[1,4]{Trevor Garrity}
\author[1]{Wyatt Pochman}

\affil[1]{Brigham Young University}
\affil[2]{Bredesen Center, University of Tennessee}
\affil[3]{University of Utah}
\affil[4]{University of Maryland}

\begin{document}

\maketitle

\begin{abstract}
Understanding the dynamics of the spread of diseases within populations is critical for effective public health interventions. We extend the classical SIR model by incorporating additional complexities such as the introduction of a cure and migration between cities. Our framework leverages a system of differential equations to simulate disease transmission across a network of interconnected cities, capturing more realistic patterns. We present theoretical results on the convergence of population sizes in the migration framework (in the absence of deaths). We also run numerical simulations to understand how the timing of the introduction of the cure affects mortality rates. Our numerical results explain how localized interventions affect the spread of the disease across cities. In summary, this work advances the modeling of epidemics to a more local scope, offering a more expressive tool for epidemiological research and public health planning.
\end{abstract}

\section{Introduction}
\label{sec:introduction}
Disease transmission is often modeled using a contact graph in which nodes represent individuals and edges represent contact between the individuals (\cite{toroczkai2007proximity}, \cite{leung2020contact}). Two main approaches can then be taken when utilizing this contact graph:
\begin{enumerate}
\item Use the fine-grained contact model of the disease, giving us information about specific individuals or;
\item Ignore the graph topology and instead report aggregate information on the spread of the disease over time by solving ODEs.
\end{enumerate}
Note that the first approach has been explored in other works such as ~\cite{keeling}, which investigates how individual-level changes to mixing patterns impact the speed and dynamics of disease spread in a population.

We focus on the second approach in this work, but we still eye the appeal of the first and seek implement it on a ``macro" scale. Our goal is to mitigate a fundamental limitation of traditional epidemic ODE models: they tell us nothing about \emph{where} the disease that is most prevalent within a population. That is, most ODE epidemic models report proportions of people (i.e. susceptible, infected, and recovered) over the \emph{entire} population. By instead representing the population as several (connected) cities, we can construct an ODE model that gives information about the disease within each city locally. This type of problem has been researched before; for example, ~\cite{multi_group} investigate global stability of multi-group epidemic models where the groups are permitted to move between different populations. We pose a similar model and expand upon it by adding dynamics of a disease cure (or vaccine).

\section{Background}

In epidemiology, there are several basic models commonly considered in the literature.
\cite{Brauer2008CompartmentalMI} give an overview of the SIR, SIS, SEIS, and SIRS models, each of which ``compartmentalizes" the population into different distinct groups. In each modeling scenario, treatment and quarantine measures need to be applied to the correct group(s) of individuals in order to optimally slow the spread of the disease. The SIR model, the most basic of these four models, is a fundamental building block of our proposed multi-city model. The SIR model splits up a population into three groups (susceptible, infected, and recovered) and makes the following modeling choices:
\begin{itemize}
    \item The proportion of susceptible individuals decreases at a rate proportional to the product of the infected and susceptible populations;
    \item The proportion of infected individuals grows at the above rate (as more people become infected) and decreases at a rate proportional to the infected population (as people recover);
    \item The proportion of recovered individuals grows as the infected individuals recover.
\end{itemize}
We can also introduce the effects of a cure into the SIR model. \cite{beckley2013modeling} modeled the effect of a cure on disease spread, a model we also explored. They found that for the SIR model
\begin{equation}
\begin{aligned}
    \dot{S} &= (1-p)m - \beta SI - mS\\
    \dot{I} &= \beta SI - \gamma I - mI\\
    \dot{R} &= pm + \gamma I - mR
\end{aligned}
\end{equation}
where $m$ now represents the birth/death rate and $p$ represents the percent of newborns who are born with a cure (and hence effectively born recovered). There are two equilibria: One is at $(S, I, R) = (1-p, 0, p)$ and is known as the disease free equilibrium (DFE). The other is at
\begin{equation}
(S, I, R) = \left(\frac{1}{R_0}, \frac{mR_0(1-p) - m}{\beta}, p + \frac{\gamma(R_0(1-p) - 1)}{\beta} \right)
\end{equation}
where $R_0 = \frac{\beta}{\gamma + m}$. This is known as the Endemic Equilibrium (EE). 

A simple analysis of the eigenvalues of the linearization at the equilibria shows that, when $R_0(1-p)<1$, the DFE is stable and the EE is unstable. However, when when $R_0(1-p) > 1$, the DFE is unstable and the EE is stable . Thus, there is a bifurcation at $R_0(1-p) = 1$ or $p = 1 - \frac{1}{R_0}$. When the percent of newborns born with the cure (or vaccinated) is greater than this value, the population will converge to the DFE, meaning the disease is effectively eradicated. Conversely, if the percent of newborns born with the cure is less than this value, then the population will converge to the EE, meaning the disease will never be eradicated, and a constant proportion of the population will be infected. Note that since there is at most one stable equilibrium, there will never be a hysteresis, as the population will always converge to the stable solution regardless of initial values (assuming the initial infected population is not exactly 0).

\section{Modeling}

Building upon the conventional SIR model, we iteratively propose several variants that incorporate additional epidemiological dynamics. These extensions address key limitations of the basic model by integrating mechanisms for a cure and for interactions across interconnected population hubs. Section \ref{Final_Model} describes the final model, which integrates all of our proposed modifications.

\subsection{Multi-City Migration Model}
\label{Migration_Model}
\begin{figure}[htb]
\begin{center}
\includegraphics[width=.30\linewidth]{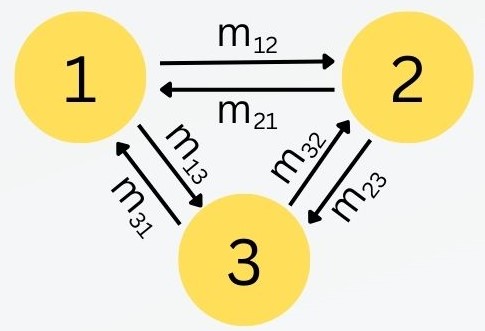} 
\end{center}
\caption{The migration model with three cities. Each population has incoming and outgoing migrants. $m_{12}$, for example, denotes the rate of migration from the first to second population.}
\label{fig:SimpleGraph}
\end{figure}

We wish to extend the basic SIR model to the case where a population is made up of multiple ``hubs" or cities. We start with a simple model that simulates the population changes between three cities (Figure \ref{fig:SimpleGraph}) (without yet accounting for the spread of a disease). This model makes these basic assumptions:
\begin{itemize}
    \item Each city sends emigrants to every other city at rates proportional to their populations.
    \item The total population of all three cities combined is constant; there are no deaths or births.
    \item The populations of each city are large enough that we can approximate the discrete changes with a continuous model.
\end{itemize} 

First, consider the population, $P_1$, of the first city. The rates at which emigrants leave for other cities is proportional to the first city's population. Simultaneously, we have incoming immigrants from the other cities. So, we have $\frac{dP_1}{dt}=m_{21}P_2+m_{31}P_3-(m_{12}+m_{13})P_1$ (where each $m_{ij}\geq 0$ is a hyperparameter). Following this process for the other two populations gives the system of equations (in matrix form):
\begin{equation}
    \frac{d}{dt}\begin{bmatrix}P_1\\P_2\\P_3\end{bmatrix}=\begin{bmatrix}-m_{12}-m_{13}&m_{21}&m_{31}\\m_{12}&-m_{21}-m_{23}&m_{32}\\m_{13}&m_{23}&-m_{31}-m_{32}\end{bmatrix}\begin{bmatrix}P_1\\P_2\\P_3\end{bmatrix}
\end{equation}\label{MigrationModel3}

We can also generalize this to an arbitrary fully connected model of $n$ cities with the ODE $\dot{P}=MP$
\begin{equation}
    \label{MigrationModelN}
    \frac{d}{dt}\begin{bmatrix}P_1\\P_2\\\vdots\\P_n\end{bmatrix}=\begin{bmatrix}-\sum_{i\not=1}m_{1i}&m_{21}& \dots &m_{n1}\\m_{12}&-\sum_{i\not=2}m_{2i}&\dots &m_{n2}\\\vdots&\vdots&\ddots&\vdots\\m_{1n}&m_{2n}&\dots&-\sum_{i\not=n}m_{ni}\end{bmatrix}\begin{bmatrix}P_1\\P_2\\\vdots\\P_n\end{bmatrix}
\end{equation}
Note that the sum of each of the columns is $0$, the values on the diagonal are nonpositive, and the off-diagonal entries are nonnegative. A more thorough analysis of this model is provided in Section \ref{Steady_State_City}

\subsection{Multi-City SIR Model}
\label{multiSIR}
We now expand the migration model to include an independent SIR model in each city; that is, we assume that a disease may only spread between members of the same city. This is a better model of real-world population dynamics than the conventional SIR model, which assumes a given individual is equally likely to receive or transmit the disease from all other individuals. To simplify the analysis, we assume the state of infection does not impact the migration of individuals. The ODE for any given population $i$ is then:
\begin{equation}\label{Multi-Pop SIR Model}
\begin{aligned}
    \dot{S}_i&=-\sum_{k\not=i}m_{ik}S_{i} +\sum_{k\not=i}m_{ki}S_k- \beta S_{i}I_i\\
    \dot{I}_i&=-\sum_{k\not=i}m_{ik}I_{i} +\sum_{k\not=i}m_{ki}I_k+\beta S_{i}I_i-\gamma_i I_i\\
    \dot{R}_i&=-\sum_{k\not=i}m_{ik}R_{i} +\sum_{k\not=i}m_{ki}R_k+\gamma I_i
\end{aligned}
\end{equation}

Note that $S_i$, $I_i$ and $R_i$ represent the proportion of susceptible, infected, and recovered people in population $i$ relative to the total initial population. In this model, the population of each city is not guaranteed to be zero, but since $\sum_{i=1}^n (\dot{S} +\dot{I}+\dot{R}) = 0$ the total population will remain constant. Results for this model are included in Appendix \ref{appendix:c}.

\subsection{Single-City SIR with Cure}
\label{SinglecityCure}
Unlike the standard SIR model, we are also interested in modeling the effectiveness of introducing a cure. To do this, we replace the recovered population with a dead population and now only allow the recovered population to grow after a fixed time $t_c$ in which a cure is introduced. More specifically,
\[
\dot{S} = -\beta SI,\quad \dot{I} = \beta SI - c_t I - \gamma I,\quad \dot{D} = \gamma I,\quad \dot{R} = c_t I
\]
where $c_t$ is equal to $0$ for $t < t_c$ and is equal to $c$ for $t \geq t_c$.

\subsection{Multi-City SIR with Cure}
\label{multi_city}
Just as in the single-city case, we can adapt the multi-city SIR model to include the possibility of a cure:
\label{Final_Model}

\begin{equation}
\label{Multi-city-cure}
\begin{aligned}
        \dot{S}_i &= -\beta S_i I_i - \sum_{j \neq i} m_{ij} S_i + \sum_{j \neq i} m_{ji} S_j\\
        \dot{I}_i &= \beta S_i I_i - \gamma I_i - C_i I_i - \sum_{j \neq i} m_{ij} I_i + \sum_{j \neq i} m_{ji} I_j\\
        \dot{R}_i &= C_i I_i - \sum_{j \neq i} m_{ij} R_i + \sum_{j \neq i} m_{ji} R_j\\
        \dot{D}_i &= \gamma I_i\\
        \dot{C}_i &= c_0 \delta(t-t_c) +\sum_{j \neq i} m_{ji}  C_j - \sum_{j \neq i} m_{ij}  C_i
\end{aligned}
\end{equation}
where $S_i$, $I_i$, $R_i$, $D_i$, and $C_i$ are the number of susceptible, infected, recovered, dead, and amount of cure in region $i$ respectively. $\beta$ is the transmission rate, $\gamma$ is the fatality rate, $m_{ij}$ is the rate of movement from region $i$ to region $j$, and $c_0$ is the amount of cure is introduced to each city at time $t_c$. Note that the amount of cure $C_i$ starts at 0 for each $i$.\\

\section{Results}

In the following subsections we explore the behavior of our different population/disease spread models both theoretically and quantitatively.

\subsection{Steady State of the Multi-City Migration Model}
\label{Steady_State_City}

\subsubsection{Numerical Approximation}

\begin{figure}[ht]
\label{fig:migration-2}
\begin{subfigure}[t]{0.45\textwidth}
    \centering
    \includegraphics[width=1\textwidth]{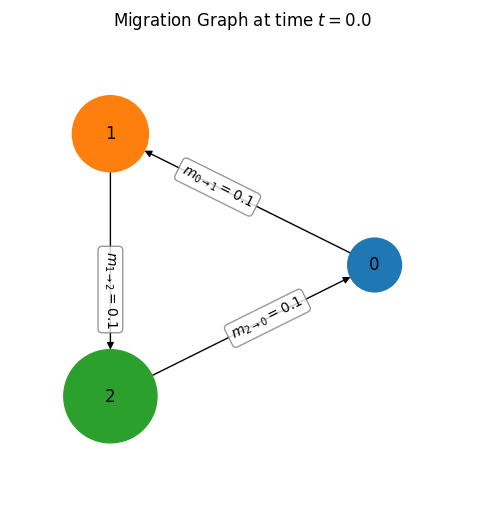}
    \caption{A network of three cities, with migration occurring in a counterclockwise fashion. The node radii represent the relative initial sizes of each population.}
\label{fig:migration-example}
    \end{subfigure}
    \hfill
\begin{subfigure}[t]{0.45\textwidth}
    \centering
    \includegraphics[width=1\textwidth]{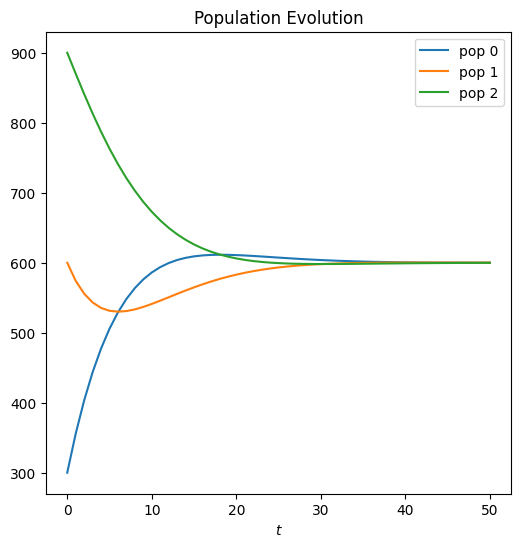}
    \caption{The evolution of the populations in the triangular system as time elapses. The initial populations are 300, 600, and 900. The populations in each city tend to the stable equilibrium 600 over time.}
\label{fig:migration-flow}
        \end{subfigure}
\end{figure}

We numerically approximate the solution to the model from Section \ref{Migration_Model} with three cities. Figures \ref{fig:migration-example} and \ref{fig:migration-flow} demonstrate how population sizes fluctuate over time in a triangular network. Migration flow occurs steadily between the three populations so the solution tends to a fixed point as $t\to\infty$.

\subsubsection{Theoretical Results}
Here we prove that the matrix $M$ in Equation \ref{MigrationModelN} is uniformly stable and that a population will hence converge to its projection onto the 0-eigenspace. The population satisfies $\dot{P} = MP$ and hence $P(t) = \exp(Mt)P(0)$.

Note that the matrix $M$ is nonnegative off of the diagonal. Such matrices are known as a Metzler matrix. The columns of $M$ also sum to 0. We seek to show that $M$ is uniformly stable, meaning all eigenvalues of $M$ have negative real part except for the eigenvalue 0 which is semi-simple. Below we prove the result when $M$ is irreducible (after shifting by a constant times the identity function). Surprisingly, we could not find this result for general $M$ anywhere in the literature. We therefore provide a novel proof in Appendix \ref{appendix:metzler}, using some ideas from \cite{GrigPresent}.

To begin, we will assume $M$ is Metzler (so the off-diagonal elements are nonnegative) and will assume that the sum of each of the columns of $M$ is equal to 0. There then exists $p \geq 0$ such that $M + pI$ is nonnegative. Since the sum of any column of $M$ is equal to 0, the sum of any column of $M + pI$ will be equal to $p$. Combined with the matrix being nonnegative, this implies $\|M + pI\|_1 = p$ and hence $r(M) \leq p$. Therefore, every eigenvalue of $M+pI$ is at most distance $p$ from the origin. Therefore, every eigenvalue of $M = (M + pI) - pI$ is at most distance $p$ from $-p$ and hence will have negative real part with the possible exception of the eigenvalue 0.

Finally, the vector consisting entirely of 1's will be a left eigenvector or $M + pI$ with eigenvalue $p$. Therefore, $r(M) = p$. Since $M + pI$ is nonnegative, it follows from the Perron-Frobenius theorem that if it is also irreducible then the eigenvalue $p$ will be simple. Shifting by $-pI$, it then follows that the eigenvalue $0 = p -p$ of $M = (M + pI) - pI$ must be simple.

More generally, we give the novel result in Appendix $\ref{appendix:metzler}$ that even when $M + pI$ is reducible, all the eigenvalues of $M$ have negative real part with the exception of 0 which is semi-simple.

Since every column of the $M$ we are considering sums to 0, it follows that 0 is an eigenvalue of $M$. By the result just proved it follows that,
\begin{align*}
\exp(Mt) &= \sum_{\lambda \in \sigma(M) } e^{\lambda t} \left(P_\lambda + \sum_{k=1}^{m_\lambda - 1} \frac{t^k D_\lambda^k}{k!} \right) \\
&= \sum_{\lambda \in \sigma(M) \setminus\{0\} } e^{\lambda t} \left(P_\lambda + \sum_{k=1}^{m_\lambda - 1} \frac{t^k D_\lambda^k}{k!} \right) + P_0.
\end{align*}
Since the zero eigenvalue is semi-simple, it does not have any $D_0$ associated with it. Furthermore all of the terms on the left have a $\lambda$ with negative real part. Thus they converge to zero as $t\rightarrow\infty$, and the whole sum approaches $P_0$.
But this is just the eigenspace associated with the 0 eigenvector. So we have:
\begin{align*}
    P(t) = \exp (M t) P(0) \rightarrow P_0 P(0)
\end{align*}
Furthermore, it is easy to see that every element of the eigenspace of 0 is an equilibrium point, as then all the other projection terms vanish. In other words, if $P(0)$ is in the eigenspace of 0 then,
\[
P(t) = \exp(Mt)P(0) = P_0 P(0) = P(0)
\]
for all $t \geq 0$. It then follows that every element of the eigenspace of 0 is a stable equilibrium. If you follow the proof that $M$ is uniformly stable even more carefully, you will find the eigenspace of 0 has a basis of positive vectors that can be scaled to sum to 1.

\subsection{Steady State of Multi-City SIR Model}
\label{SteadystateMultiSIR}

\subsubsection{Theoretical Results}

We establish theoretical results concerning the stability properties of the multi-city SIR model (Section \ref{multiSIR}). Writing out the linearization of the multi-city SIR model (equation \ref{Multi-Pop SIR Model}) gives the block matrix:
\begin{equation}
L = \begin{bmatrix}
    A_1 & M_{21} & M_{31} \\
    M_{12} & A_2 & M_{32} \\
    M_{13} & M_{23} & A_3
\end{bmatrix}
\end{equation}
where
\begin{equation}
A_i=\begin{bmatrix}
    -\beta_iI_i-\displaystyle\sum_{k=1, k\not=i}^nm_{ik}&-\beta_iS_i&0\\
    \beta_iI_i&\beta_iS_i-\gamma_i-\displaystyle\sum_{k=1, k\not=i}^nm_{ik}&0\\0&\gamma_i&-\displaystyle\sum_{k=1, k\not=i}^nm_{ik}
\end{bmatrix}
\end{equation}
and
\begin{equation}
M_{ij}=\begin{bmatrix}
    m_{ij} & 0 & 0 \\
    0 & m_{ij} & 0 \\
    0 & 0 & m_{ij}
\end{bmatrix}
\end{equation}
Notice that since this models the population between 3 cities, where each person can be susceptible, infected, or recovered, we have $M_{ij}\in \mathbb{R}^{3\times 3}$, $A_{i}\in \mathbb{R}^{3\times 3}$, and $L\in \mathbb{R}^{9\times 9}$.
Because the columns of $L$ sum to zero, total population is conserved across cities, leading to a neutral mode corresponding to the eigenvalue $0$.

To understand the other eigenvalues of $L$, it is more informative to consider $L^T$.  When all cities share identical epidemiological parameters, the coupling terms do not affect the fundamental stability condition. We make this assumption temporarily to simply the analysis of L. In this case, the global dynamics reduce to those of a single SIR system. Assume for the moment that all of the SIR models have the same hyperparameters (i.e. $\beta_i=\beta_j$ and $\gamma_i=\gamma_j$ for all $i,j$). As stated previously it is clear that zero is an eigenvalue of this matrix because $L\mathbf{0}=0$. Consider the matrices $A_i$. If we remove all of the transition probabilities $m_{ik}$ (call the new matrix $A$), then these are all the same matrix and have the same eigenvectors. Call  any such such eigenvector $v=\begin{bmatrix}
    a&b&c
\end{bmatrix}^T$. 

Notice the vector $\hat v=\begin{bmatrix}
    a&b&c&a&b&c&a&b&c
\end{bmatrix}^T$ is  an eigenvector of $L$ because all of the $m_{ik}$ will cancel out. This shows that inter-city migration preserves the basic eigenstructure of the single-city model under symmetric coupling. As a result each eigenvector of $A_i$ gives rise to the corresponding eigenvector $\hat v$ of $L^T$. Because each $A_i$ contributes 3 eigenvectors, these vectors span the eigenspace of $L^T$. Therefore it suffices to determine the eigenvalues of $A$

It is known that the eigenvalues of the standard SIR model are given by $\lambda = 0,0,\beta S-\gamma$. Now, if we introduce different parameters for each SIR model we suspect that the overall behavior of the system should still be related to the eigenvalues of each SIR model individually. Thus, the stability of the multi-city system depends on whether infection decreases in each city—roughly, when $S_i < \gamma_i / \beta_i$. 

So if we track when $S_i<\frac{\gamma_i}{\beta_i}$ that should heuristically tell us when: (1) the infected population starts decreasing and (2) when the population starts trending toward stability. As a hypothesis, since we are constantly mixing populations, we estimate that the whole system will trend toward stability once all $S_i<\frac{\gamma_i}{\beta_i}$. In other words, migration redistributes infections but does not change the overall stability threshold; the epidemic subsides once all cities pass their local thresholds.

Note that regardless of the parameters of the SIR model, we should expect that the population dynamics of the cities as a whole should behave the exact same as in the previous section. This is because the macro scale equations $\dot S_i+\dot I_i+\dot R_i=\dot P_i$ just add to the equations of the migration model. Hence, the conservation of total population ensures that coupling affects the spatial distribution of infections but not the aggregate epidemic trajectory. This result mirrors the intuition that migration redistributes individuals without altering total epidemic dynamics.

\subsection{Single-City SIR with Cure} 

\subsubsection{Numerical Approximation}
We analyze the dynamics of the single-city version of the SIR model with cure model (Section \ref{SinglecityCure})
Due non-linear nature of the ODE, we thought it best to explore its behavior numerically.

\begin{figure}[ht]
    \centering
    \begin{subfigure}{0.4\textwidth} 
        \centering
        \includegraphics[width=\textwidth]{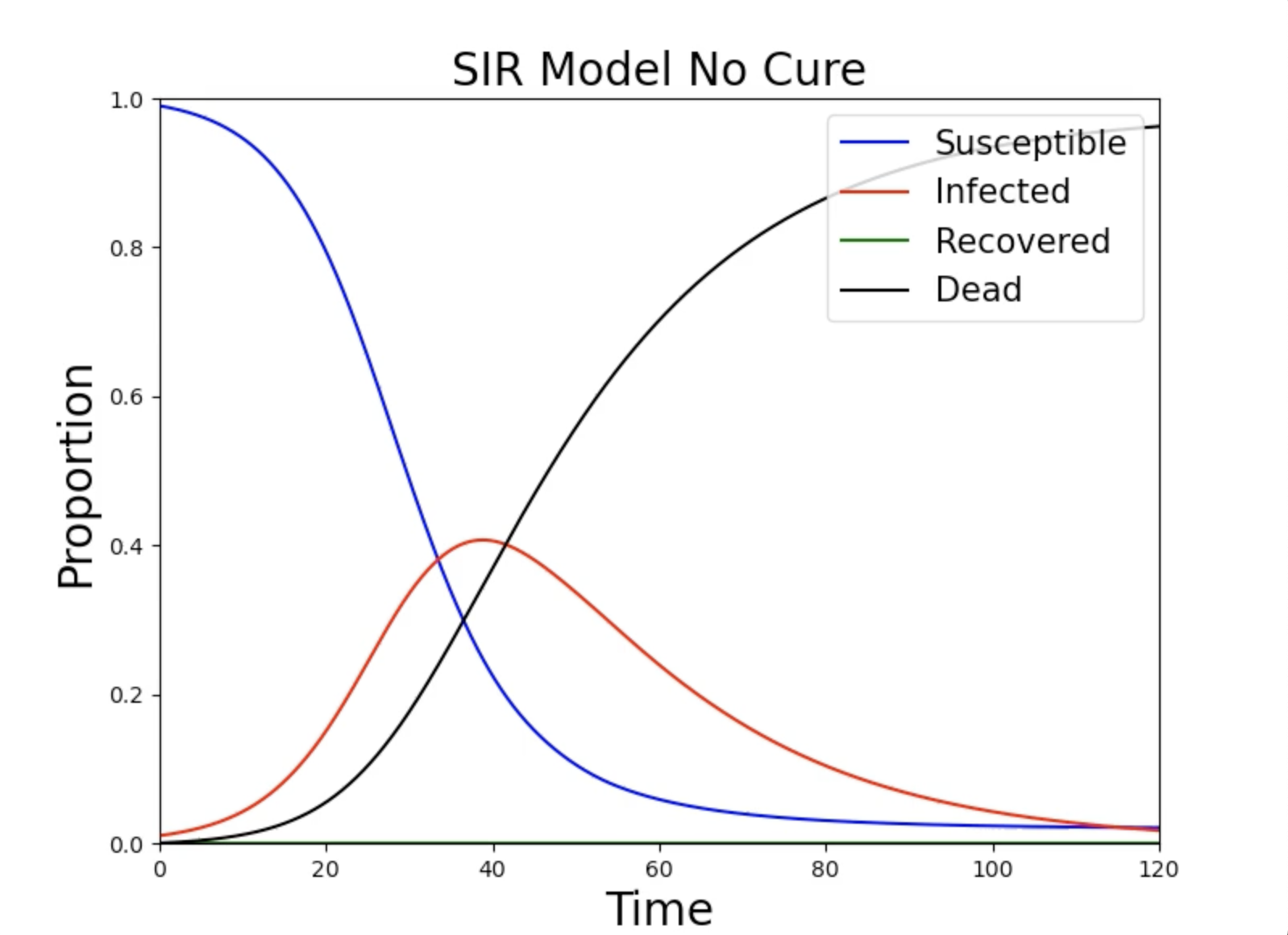}
        \caption{This plot serves as a baseline, showing the dynamics in the absence of any cure intervention. The infected population gradually increases and then decreases to $0$.}
        \label{fig:no_cure}
    \end{subfigure}
    \hfill
    \begin{subfigure}{0.4\textwidth} 
        \centering
        \includegraphics[width=\textwidth]{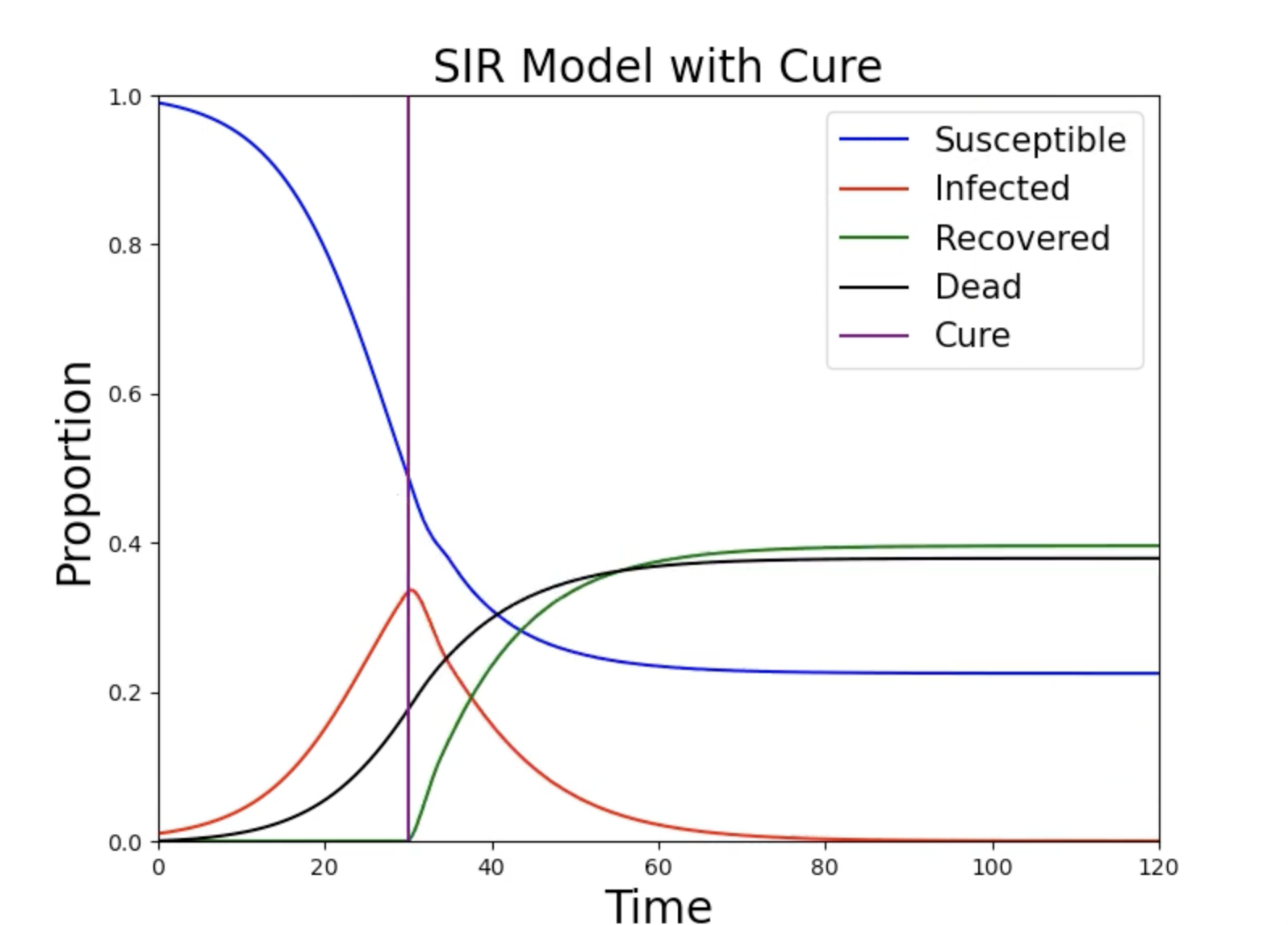}
        \caption{Here, we model the introduction of a cure early in the outbreak, demonstrating its significant effect in reducing the number of infections and fatalities.}
        \label{fig:early_cure}
    \end{subfigure}
    
    \vspace{1em} 
    \begin{subfigure}{0.4\textwidth} 
        \centering
        \includegraphics[width=\textwidth]{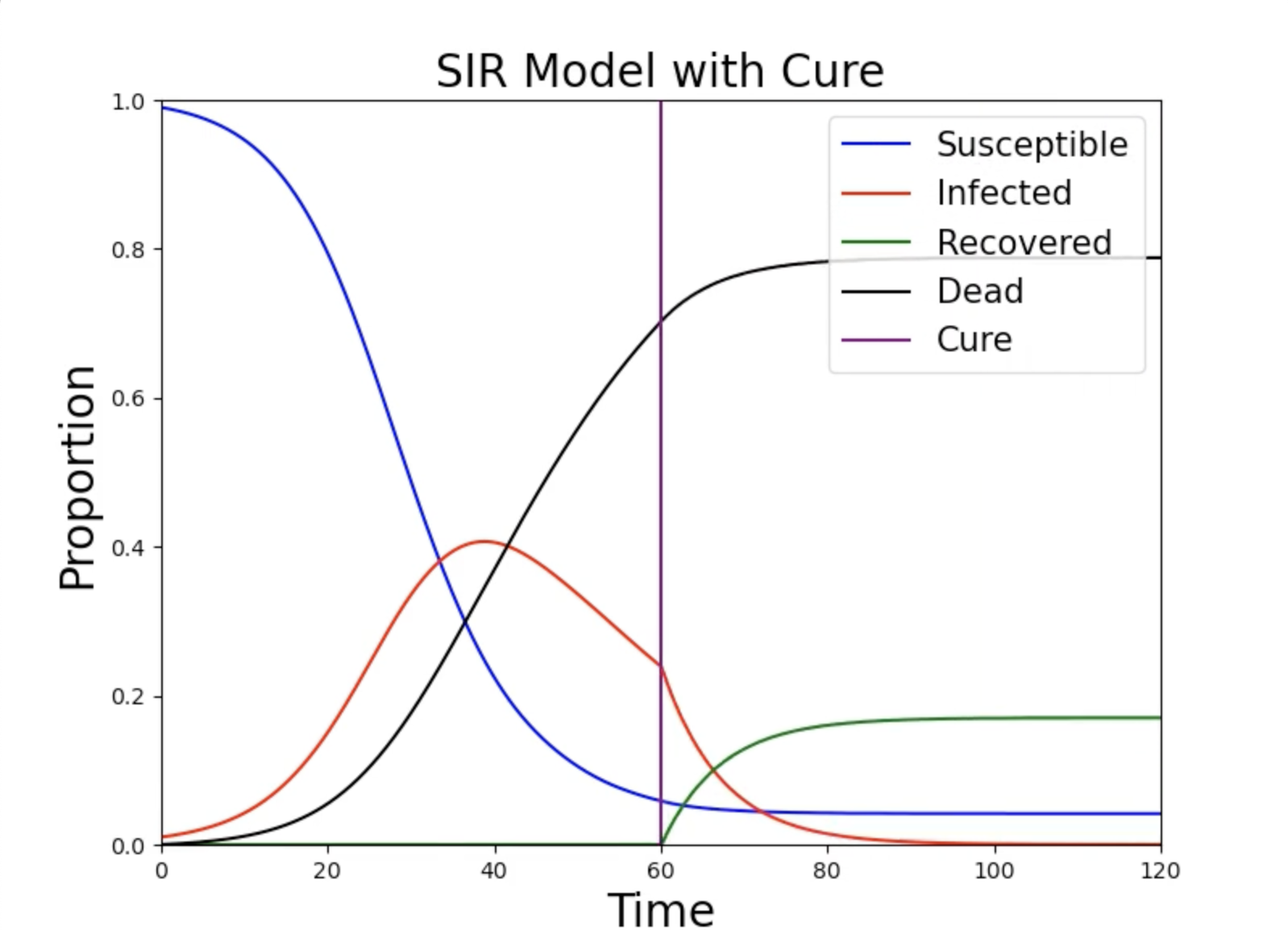}
        \caption{This scenario examines the impact of implementing a cure halfway through the outbreak. At that moment, there is a sharp decrease in the infected population.}
        \label{fig:mid_cure}
    \end{subfigure}
    \hfill
    \begin{subfigure}{0.4\textwidth} 
        \centering
        \includegraphics[width=\textwidth]{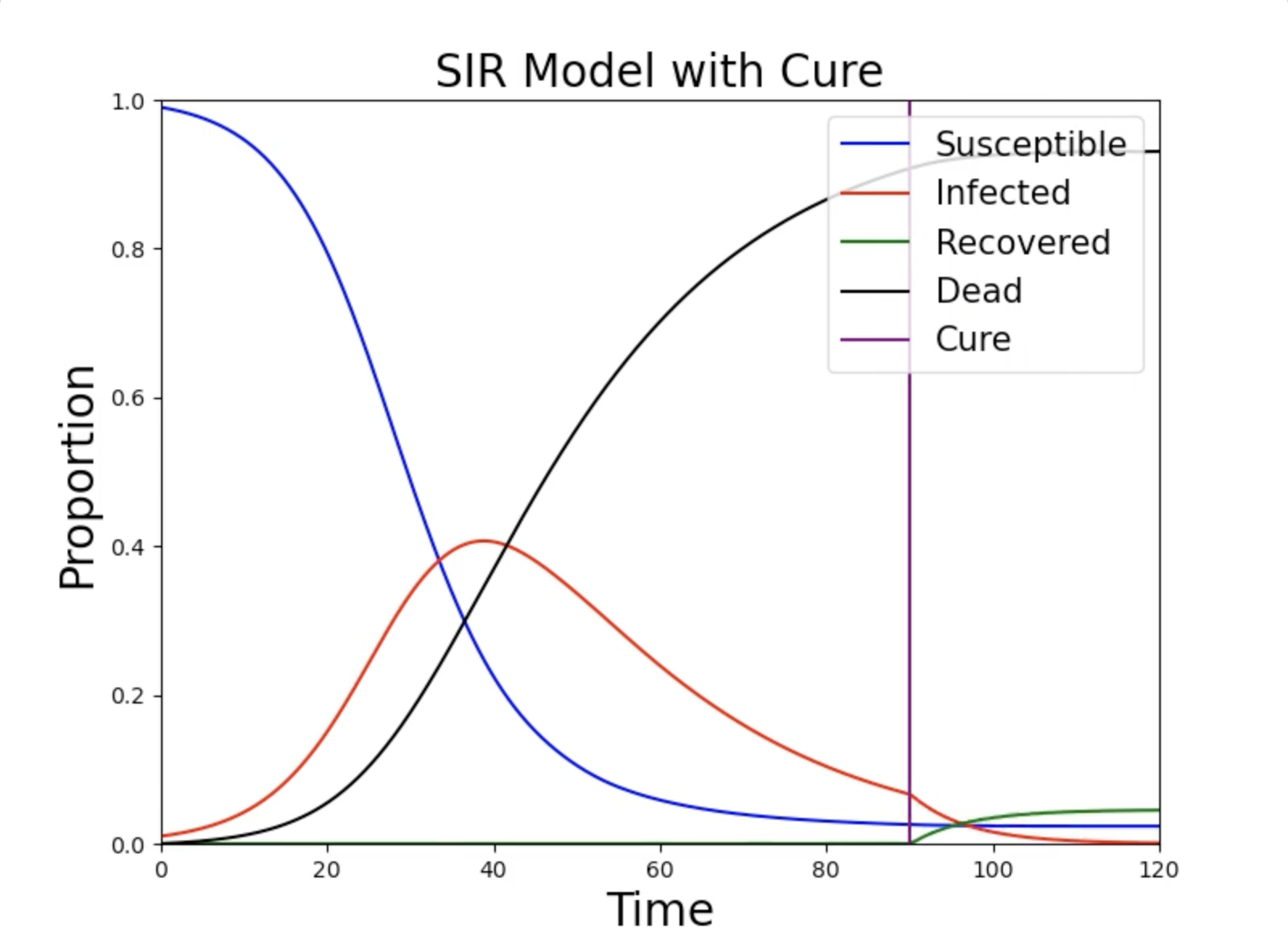}
        \caption{Finally, this plot shows the consequences of introducing a cure late in the outbreak, with limited impact due to the substantial number of fatalities.}
        \label{fig:late_cure}
    \end{subfigure}

    \caption{Comparison of different cure intervention scenarios during an outbreak. Notice the sharp change in the slope of the infected and recovered populations.}
    \label{4SIR_Plots}
\end{figure}

Using the fourth-order Runge-Kutta method, we numerically approximate the solution to the system of differential equations (Figure \ref{4SIR_Plots}).  Our parameters are given by:
\[\beta=.2,\quad \gamma=.05,\quad S(0)=.99,\quad I(0)=.01,\quad R(0)=0. \quad D(0)=0\]
We found that, consistent with our intuition, introducing a cure earlier results in fewer deaths as people recover quicker get infected less. Interestingly, there seems to be some form of shock introduced into the model when the cure is introduced (evidenced by the sharp change in slope in the red curves of Figure \ref{4SIR_Plots})

\subsection{Multi-City SIR with Cure}

\subsubsection{Numerical Approximation}

Again, using the fourth-order Runge-Kutta method, we approximate the solution to this expanded system of differential equations (Equation \ref{Multi-city-cure} from Section \ref{multi_city}). This method enables us to model the dynamics of disease spread across multiple regions, incorporating the effects of mobility between regions and the introduction of a cure at different stages. For our simulations, we assume $\beta=\beta_i$ and $\gamma=\gamma_i$ for all $i$; that is, the $\gamma$ and $\beta$ parameters for each population are constant.\\

\begin{figure}[ht]

\begin{subfigure}[t]{0.45\textwidth}
    \centering
    \includegraphics[width=\textwidth]{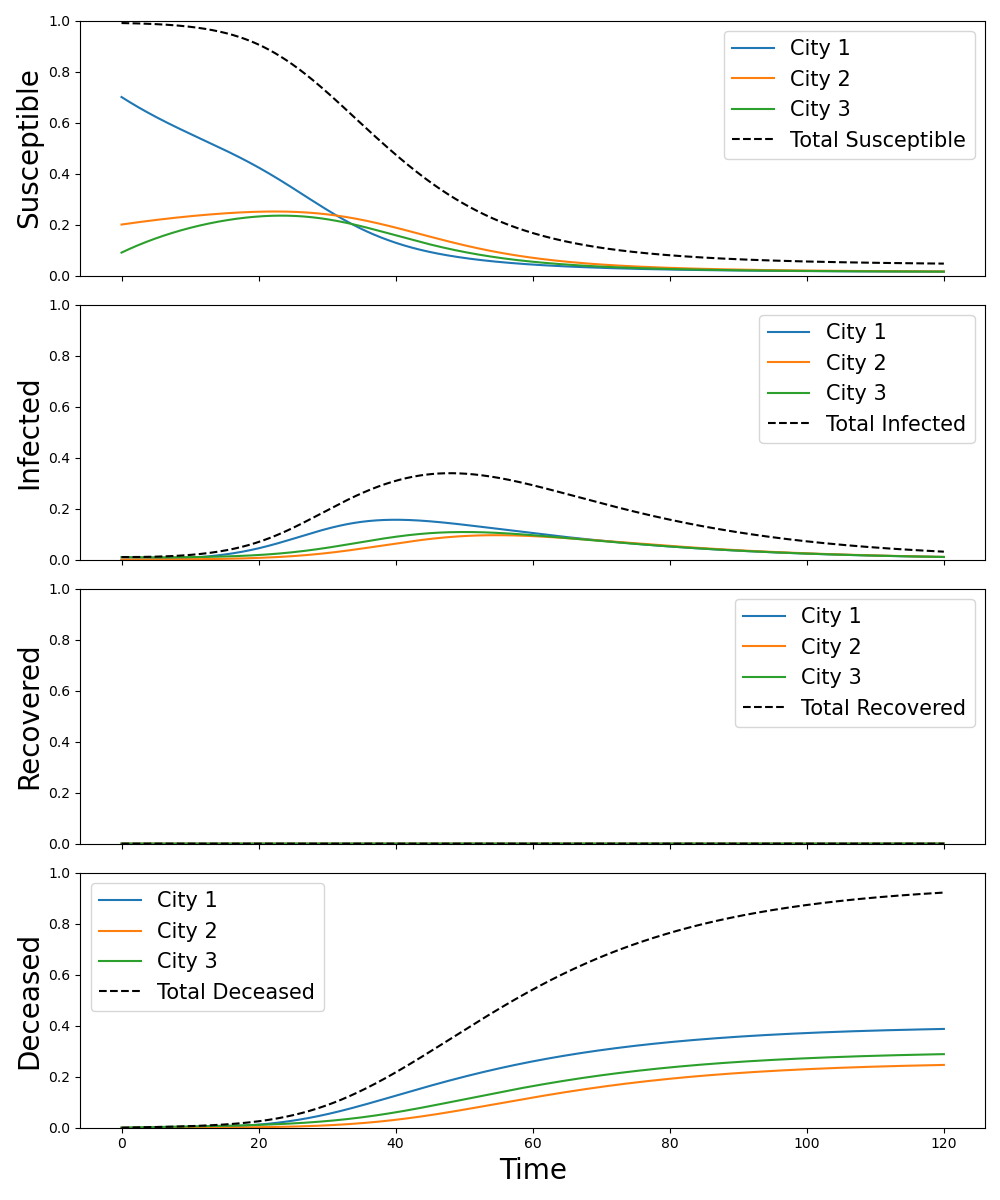}
    \caption{This plot serves as a baseline, depicting the spread of disease in a multi-city network without any cure intervention. For more results, see Appendix \ref{appendix:c}}
\label{fig:multi-city-basline}
    \end{subfigure}
    \hfill
\begin{subfigure}[t]{0.45\textwidth}
    \centering
    \includegraphics[width=\textwidth]{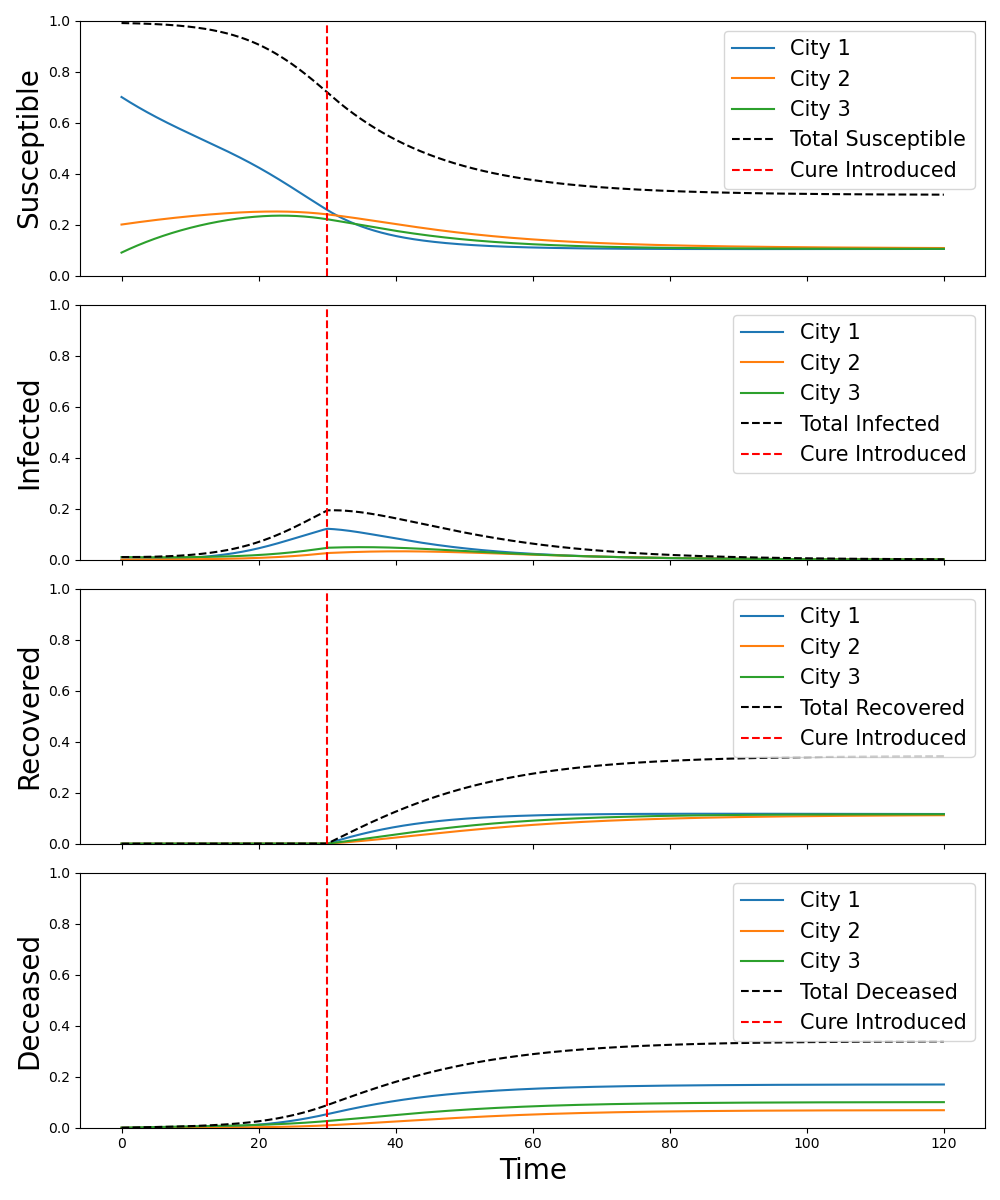}
    \caption{This scenario models the introduction of a cure early in the outbreak. The results show a substantial reduction in both infections and fatalities, with the disease largely contained before it spreads extensively across regions.}
\label{fig:multi-city-early}
\end{subfigure}
\caption{Numerical approximation for the Multi-City SIR model with and without a Cure}
\end{figure}

The plots ~\ref{fig:multi-city-basline},~\ref{fig:multi-city-early},~\ref{fig:multi-city-mid} and~\ref{fig:multi-city-late} (Appendix \ref{appendix:c}) illustrate the results of our simulations under various scenarios, comparing the impacts of cure timing on the progression of the disease in a multi-city context.

Similar to the single city case, there is a point of shock that happens when the cure is introduced into a city in the multi-city case. Interestingly, although there is change, the cure does not seem to shock cities other than the one it is introduced to. Curves in the other cities are altered more smoothly.

\section{Conclusion}

Here we discuss the appropriateness of our model and compare the results of our simulations to their expected outcomes.

\subsection{Appropriateness of Modeling} Our model includes several hyperparameters, such as a transmission rate, fatality rate, and $O(n^2)$ migration rates in the total number of cities $n$. We have inserted several simplifying assumptions, such as making the transmission and fatality rates $\beta$ and $\gamma$ consistent in each city, partially because we are interested in modeling semi-local disease spread. This model would be most appropriately applied to analyze disease spread between several large cities (like Austin, Houston, and San Antonio) rather than several states (like Tennessee, Mississippi, and Alabama). Perhaps the conditions in separate states are different enough to warrant unique parameters $\beta_i$ and $\gamma_i$, but we would expect patterns of disease spread in adjacent cities to be fairly similar.

There are several drawbacks to the approach we have taken:
\begin{itemize}
    \item The model does not prohibit infected persons from migrating from one city to the other; it assumes they migrate at the same rates as everyone else. This is done to facilitate spread of the disease; without it, cities with 0 infected persons at $t=0$ would never develop any infected persons. What actually happens in real life is probably somewhere in the middle of the two extremes. That is, infected persons will transmit the disease to other city/locations, but perhaps only at the beginning stages of the infected period when they are asymptomatic. Addressing these concerns more carefully would require an additional parameter giving the rate of migration of infected individuals, thereby complicating the model, but we have generally opted for simplicity over realism.
    \item The disease we have chosen to model with a cure is binary---it yields a 100\% fatality rate unless a cure is administered. Most diseases are not as black and white as this, and there are good reasons experts care about reducing the spread of common illnesses. For example, employees who get sick with the flu may be unable to work for several days, and if the flu propagates rapidly through a population, this could be damaging, especially within the public education system.
\end{itemize}

We note that development of a proper epidemiological model almost always requires data assimilation to verify the model is reasonable. Because the scenarios we have chosen to model are theoretical, and exact hyperparameter selection is outside the scope of this work, we cannot necessarily conclude our model is accurate at this time, but the trends produced from our simulations are reasonable enough.

\subsection{Quality of Outcomes}
Despite the limitations explained previously, our model does well at explaining why time is of the essence when developing a cure for a disease. There are definitely time delays between the point a cure is introduced into a population and the subsequent decline of the infected populace. Unfortunately, in our experiments we were unable to see any oscillatory behavior prior to the model stabilizing. However, stable equilibria are typically expected in the context of epidemiology.  For example, as expected, in a setting without a cure to a terminal disease, the entire population will eventually die.

Given more time, we plan to evaluate our model by testing it on real-world scenarios. For example, we aim to analyze data on the spread of COVID-19 across the USA, focusing on how the pandemic evolved between cities. This will allow us to assess how well our model’s predictions align with observed rates of transmission.

\section*{Acknowledgments}

We thank Dr. Tyler Jarvis and the ACME program from Brigham Young University for supporting this research project.

\bibliographystyle{plain}
\bibliography{refs}

\begin{thebibliography}{1}

\bibitem{beckley2013modeling}
Ross Beckley, Cametria Weatherspoon, Michael Alexander, Marissa Chandler, Anthony Johnson, and Ghan~S Bhatt.
\newblock Modeling epidemics with differential equations.
\newblock {\em Tennessee State University Internal Report}, 2013.

\bibitem{Brauer2008CompartmentalMI}
Fred Brauer.
\newblock Compartmental models in epidemiology.
\newblock {\em Mathematical Epidemiology}, 1945:19 -- 79, 2008.

\bibitem{multi_group}
Qianqian Cui.
\newblock Global stability of multi-group sir epidemic model with group mixing and human movement.
\newblock {\em Mathematical Biosciences and Engineering}, 16:1798--1814, 03 2019.

\bibitem{GrigPresent}
Ilie Grigorescu.
\newblock The perron frobenius theorem and a few of its many applications.
\newblock \url{https://www.math.miami.edu/~igrigore/perron-frobenius_kruczek.pdf}.
\newblock Accessed: 2024-12-06.

\bibitem{keeling}
Matt Keeling and Ken Eames.
\newblock Networks and epidemic models.
\newblock {\em Journal of the Royal Society, Interface / the Royal Society}, 2:295--307, 06 2005.

\bibitem{kermack1927contribution}
William~Ogilvy Kermack and Anderson~G McKendrick.
\newblock A contribution to the mathematical theory of epidemics.
\newblock {\em Proceedings of the royal society of london. Series A, Containing papers of a mathematical and physical character}, 115(772):700--721, 1927.

\bibitem{leung2020contact}
Abby Leung, Xiaoye Ding, Shenyang Huang, and Reihaneh Rabbany.
\newblock Contact graph epidemic modelling of covid-19 for transmission and intervention strategies.
\newblock {\em arXiv preprint arXiv:2010.03081}, 2020.

\bibitem{toroczkai2007proximity}
Zolt{\'a}n Toroczkai and Hasan Guclu.
\newblock Proximity networks and epidemics.
\newblock {\em Physica A: Statistical Mechanics and its Applications}, 378(1):68--75, 2007.

\end{thebibliography}

\appendix

\newpage
\appendix

\section{Generic SIR and Nondimensionalization}
\label{appendix:a}

To improve the stability of our analysis of the SIR model, we can perform nondimensionalization on the parameters. Since the SIR models we consider already scale the populations so that they add up to 1, the populations are already dimensionless. However, the time parameter can be nondimensionalized. 

We treat a simple SIR model of one population. The basic SIR model is well-established and can be found in \cite{kermack1927contribution}. In modern notation the SIR model is given by,
\begin{equation}
    \dot{S}=-\beta IS,\quad \dot{I}=\beta IS-\gamma I,\quad \dot{R}=\gamma I,
\end{equation}
where $S$ denotes the proportion of the population that is susceptible, $I$ denotes the proportion of the population that is infected, and $R$ denotes the proportion of the population that is recovered. Since these are proportions we set $S + I + R = 1$. Indeed we can verify $\dot{S} + \dot{I} + \dot{R} = 0$.

Since $S$, $I$, and $R$ represents proportions of the population, they are dimensionless. However, the time parameter $t$ has dimensions $[t] = \text{T}$. Therefore, the coefficients $\beta$ and $\gamma$ each have dimension $[\beta] = [\gamma] = \text{T}^{-1}$. We can replace $t$ with a dimensionless parameter $\tau$ given by $\tau = kt$ for some parameter $k$ with dimension $[k] = \text{T}^{-1}$. Then $\frac{d}{d\tau} = \frac{dt}{d\tau} \frac{d}{dt} = \frac{1}{k} \frac{d}{dt}$. If we take the derivative of $S$, $I$, and $R$ with respect to $\tau$ instead of $t$ we get
\begin{equation}
\dot{S} = -\frac{\beta}{k} IS, \quad
\dot{I} = \frac{\beta}{k} IS - \frac{\gamma}{k} I, \quad
\dot{R} = \frac{\gamma}{k} Is
\end{equation}
The two simplest choices for $k$ are $k = \beta$ and $k = \gamma$. If $k = \beta$ then,
\begin{equation}
\dot{S} = - IS, \quad
\dot{I} =  IS - \gamma' I, \quad
\dot{R} = \gamma' I
\end{equation}
where $\gamma' = \gamma/\beta$ is a dimensionless parameter. If $k = \gamma$ then,
\begin{equation}
\dot{S} = -\beta' IS, \quad
\dot{I} = \beta' IS - I, \quad
\dot{R} = I
\end{equation}
where $\beta' = \beta/\gamma$ is a dimensionless parameter. In other words, nondimensionalization allows us to replace one of the constants in the $SIR$ model with 1. We could also replace any of the populations with a dimensional constant times that population, but then the coefficients would no longer be the same across equations and the populations would no longer sum to 1.

A more general SIR model involves $n$ dimensionless population proportions $P_1, \dots, P_n$ that satisfy,
\begin{equation}
\dot{P}_i = \sum_{\{j_1, \dots, j_k\} \subset \{1, \dots, n\}} a_{i, j_1, \dots, j_k} P_{j_1} \cdots P_{j_k}
\end{equation}
for $1 \leq i \leq n$. Again, each of the populations are dimensionless and $[a_{i, j_1, \dots, j_k}] = \text{T}^{-1}$ for each coefficient $a_{i, j_1, \dots, j_k}$. Again, if $k$ is some constant with $[k] = \text{T}^{-1}$ then replacing time with $\tau = kt$ gives
\begin{equation}
\dot{P}_i = \sum_{\{j_1, \dots, j_k\} \subset \{1, \dots, n\}} \frac{a_{i, j_1, \dots, j_k}}{k} P_{j_1} \cdots P_{j_k}
\end{equation}
Just as above we can set $k$ equal to one of the coefficients in order to effectively replace one of the coefficients with 1.

\section{Analysis of General Population Matrices}
\label{appendix:metzler}
Here we prove that for a general Metzler matrix $M$ for which the sum of each column is less than or equal to 0, the matrix $M$ is uniformly stable. Equivalently, we prove that all the eigenvalues of $M$ are negative with the possible exception of 0 which must be semi-simple.

To begin, we will assume $M$ is Metzler (so the off-diagonal elements are nonnegative) and will assume that the sums of the columns of $M$ are less than or equal to 0. There then exists $p \geq 0$ such that $M + pI$ is nonnegative. Since the sum of any column of $M$ is less than or equal to 0, the sum of any column of $M + pI$ will be less than or equal to $p$. Combined with the matrix being nonnegative, this implies $\|M + pI\|_1 \leq p$ and hence $r(M) \leq p$. Therefore, all the eigenvalues of $M+pI$ are at most distance $p$ from the origin. Therefore, all the eigenvalues of $M = (M + pI) - pI$ are at most distance $p$ from $-p$ and hence all have negative real part with the possible exception of the eigenvalue 0.

If $M + pI$ is irreducible, there are two cases. First, if the sum of every column of $M$ is exactly equal to $p$, then the vector of 1's will be a left eigenvector or $M + pI$ with eigenvalue $p$. By the Perron-Frobenius theorem, it then follows that $p = r(M)$ is a simple eigenvalue of $M$, implying $0 = p - p$ is a simple eigenvalue of $M = (m + pI)-pI$. Second, if the sum of one of the columns of $M$ is less than $p$, we claim $p$ is not an eigenvalue of $M + pI$ and that $0$ is hence not an eigenvalue of $M$. For if $p$ is an eigenvalue of $M + pI$, then $r(M) = p$ and by the Perron-Frobenius theorem, $M + pI$ would have a positive eigenvector $v$ associated with the eigenvalue $p$. Let the $i$th column of $M+pI$ be denoted by $a_i$. Then since $v_i > 0$ for each $i$ and $\|a_i\| \leq p$ for each $i$ with one of the inequalities being strict we can calculate,
\begin{align*}
\|(M+pI) v\|_1 &= \|v_1 a_1 + \cdots + v_n a_n\|_1 \\
&= v_1 \|a_1\|_1 + \cdots + v_n\|a_n\|_1 \\
&< v_1 p + \cdots + v_n p\\
&= \|pv\|_1
\end{align*}
which is a contradiction as $(M+pI)v = pv$ and hence $\|(M+pI)v\|_1 = \|pv\|_1$. Therefore, $p$ is not an eigenvalue of $M + pI$ and consequently 0 is not an eigenvalue of $M$, implying all the eigenvalues of $M$ have negative real part.

Finally, suppose $M$ might not be irreducible. By what was initially shown above we know all the eigenvalues of $M$ have negative real part with the possible exception of the eigenvalue 0. It remains to prove the eigenvalue 0 has no nilpotent part or that there are no generalized eigenvectors. This is equivalent to showing $M^2v = 0$ implies $Mv = 0$ for any $v$. Note that it is well-known that any reducible matrix can be written in block upper/lower-triangular form with the block matrices along the diagonal being irreducible simply by permuting its indices appropriately. Without loss of generality, we may therefore assume $M+pI$ is of the form:
\[
M + pI = \begin{pmatrix} B_{11} & 0 & 0 \\ \vdots & \ddots & 0 \\ B_{k1} & \cdots & B_{kk} \end{pmatrix}
\]
where each block along the diagonal $B_{ii}$ is square and irreducible. Furthermore, since $M + pI$ is nonnegative off of the diagonal, the columns of each $B_{ii}$ must sum to at most $p$. It also follows that,
\[
M = \begin{pmatrix} B_{11} - pI & 0 & 0 \\ \vdots & \ddots & 0 \\ B_{k1} & \cdots & B_{kk} - pI\end{pmatrix}.
\]
 Suppose $M^2v = 0$ and let $[v]_i$ denote the $i$th block of $v$ for any $1 \leq i \leq k$. To prove $Mv = 0$ it suffices to prove $B_{ij}[v]_j = 0$ when $i > j$ and $(B_{ii}-pI)[v]_i = 0$ for each $i$. Suppose by induction $B_{jk}[v]_k = 0$ when $j > k$ and $(B_{kk}-pI)[v]_k = 0$ for all $k \leq i-1$ (starting with the trivial $i = 1$). Then $[M v]_i = \sum_{k <i } B_{ik} [v]_k + (B_{ii} - pI) [v]_i = (B_{ii} - pI)[v]_i$ by our inductive hypothesis. Since the inductive hypothesis also implies $[Mv]_k = 0$ for $k < i$, it follows that $[M^2v]_i = \sum_{k < i} B_{ik}[Mv]_k + (B_{ii} - pI) [Mv]_i = 0 + (B_{ii} - pI)^2 [v]_i$. Therefore, $0 = [M^2v]_i = (B_{ii} - pI)^2 [v]_i$. Since $B_{ii}$ is irreducible, it follows from the work above, that if $B_{ii} - pI$ has 0 as an eigenvalue, it must be simple. Therefore, $(B_{ii} - pI)^2[v]_i = 0$ implies $(B_{ii} - pI)[v]_i = 0$.
 
 It remains to show $B_{ji} [v]_i = 0$ for $j > i$. There are two cases. If $B_{ji} = 0$ for all $j > i$ then trivially $B_{ji} [v]_i = 0$ for $j > 0$. Otherwise, one of the columns $B_{ji}$ will have a positive element. Since that entire column in $M + pI$ sums to at most $p$, it follows that the corresponding column in $B_{ii}$ will sum to less than $p$. Then by the work above, $B_{ii} - pI$ will not have 0 as an eigenvalue, and $(B_{ii} - pI)^2$ will hence not have 0 as an eigenvalue. Therefore, $(B_{ii} - pI)^2[v]_i = 0$ implies $[v]_i = 0$, so once again $B_{ji} [v]_i =0$ for $j > i$. This completes the induction, showing $M^2 v = 0$ implies $Mv = 0$. If 0 is an eigenvalue of $M$, it follows that it cannot have any nilpotent part. To summarize, all the eigenvalues of $M$ must have negative real part with the possible exception of the eigenvalue 0 which must be semi-simple. Equivalently, $M$ will be uniformly stable.

\section{Multi-City SIR Plots}
\label{appendix:c}

\begin{figure}[ht]
\label{fig:multi_city-2}
\begin{subfigure}[t]{0.45\textwidth}
    \centering
    \includegraphics[width=1\textwidth]{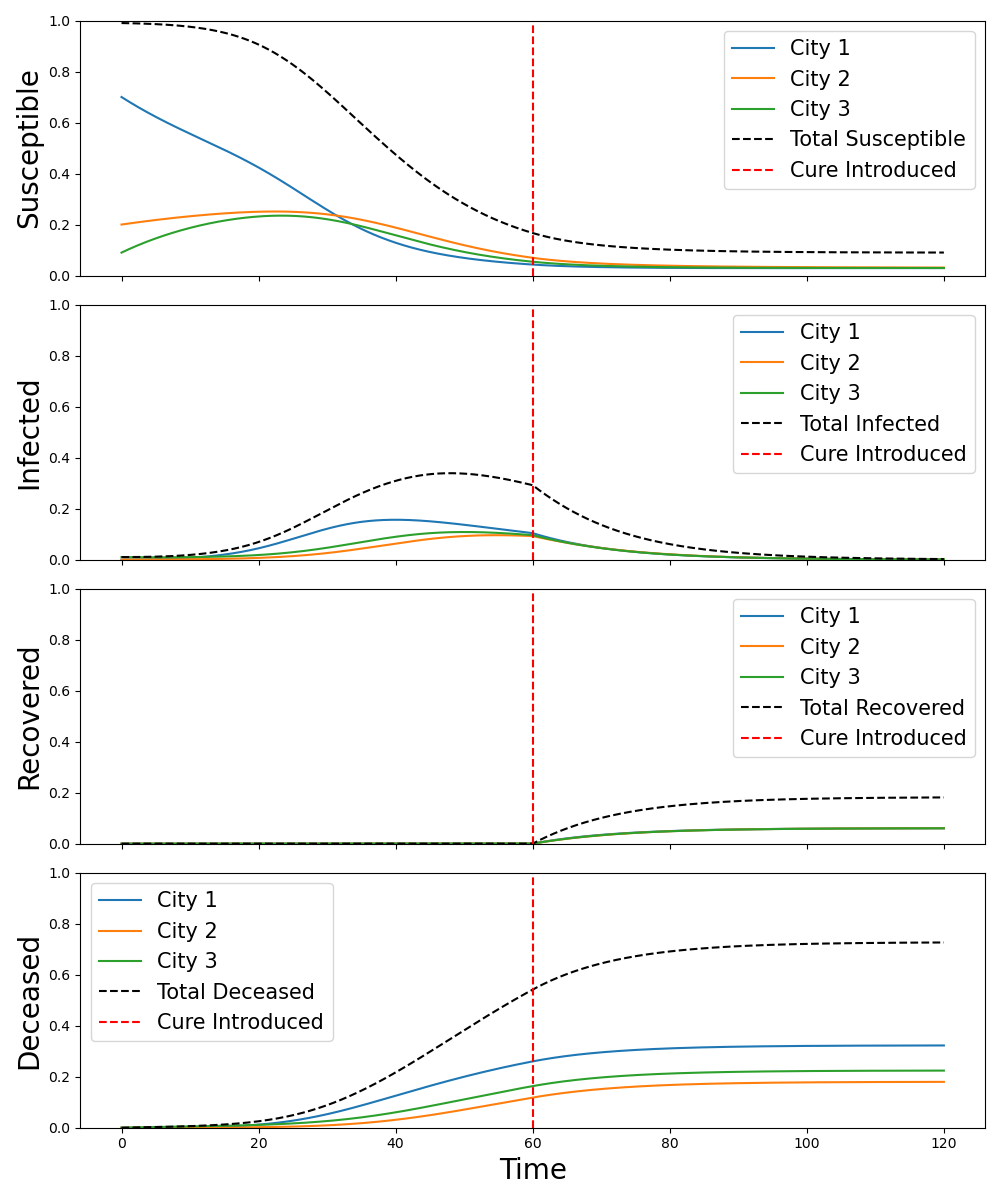}
    \caption{Here, the cure is introduced after the disease has already spread to multiple regions. While infections and fatalities are mitigated, the delay in introducing the cure results in higher peak infection rates compared to the early cure scenario.}
\label{fig:multi-city-mid}
    \end{subfigure}
    \hfill
\begin{subfigure}[t]{0.45\textwidth}
    \centering
    \includegraphics[width=1\textwidth]{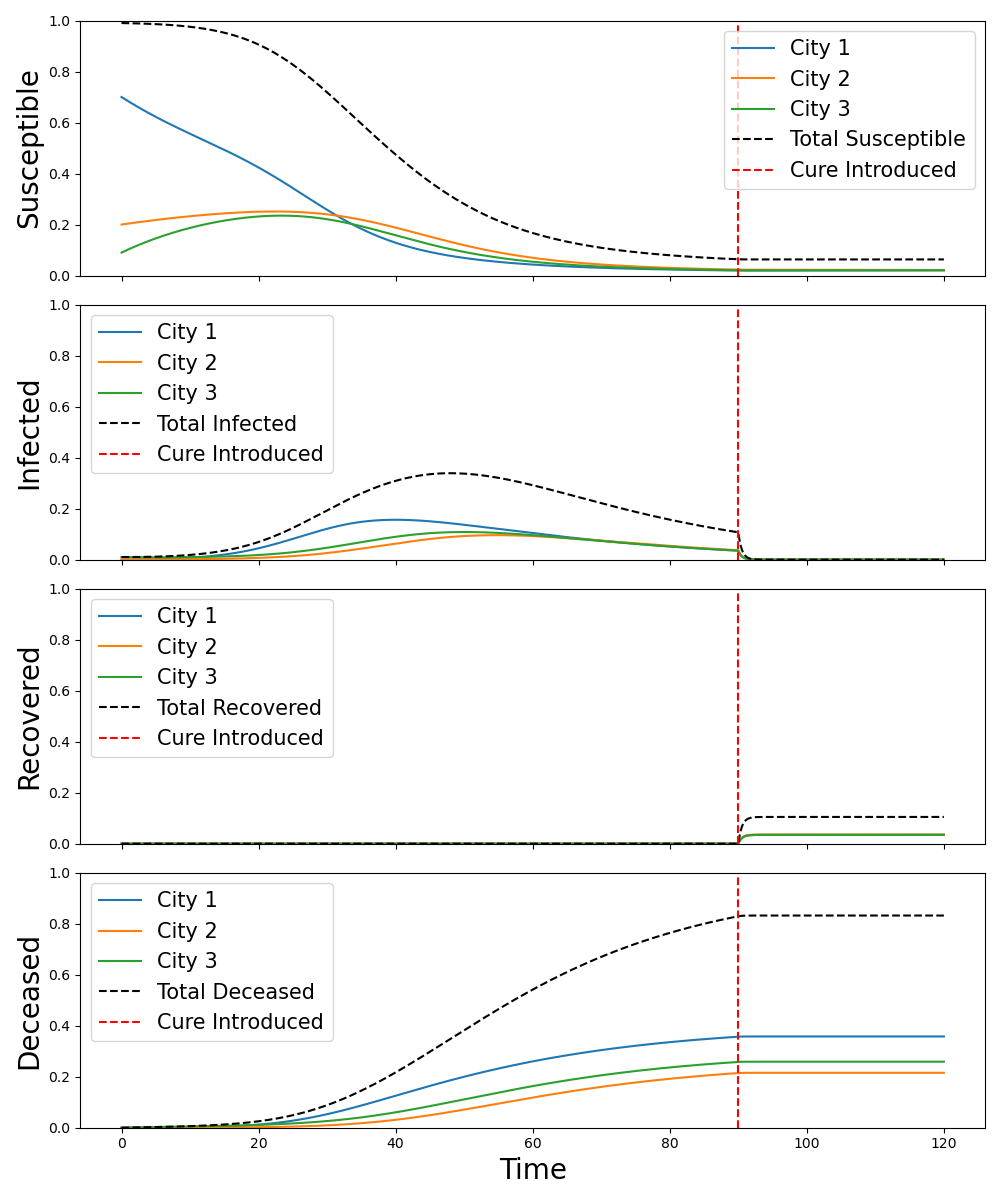}
    \caption{Finally, this plot examines the consequences of a delayed cure intervention. By this point, the disease has widely spread, and the introduction of the cure has minimal impact in reducing infections or fatalities.}
\label{fig:multi-city-late}
        \end{subfigure}
\end{figure}

\end{document}